\documentclass[11pt,twoside]{article}
\usepackage{asp2014}

\aspSuppressVolSlug
\resetcounters

\begin{document}

\title{Production, Processing and Consumption of the Dust in the Galaxy}
\author{George~Gontcharov,$^1$
\affil{$^1$Central (Pulkovo) Astronomical Observatory, Russian Academy of Sciences, 
65/1 Pulkovskoye Shosse, Saint-Petersburg, 196140 Russia; \email{georgegontcharov@yahoo.com}}}

\paperauthor{George~Gontcharov}{georgegontcharov@yahoo.com}{orcid.org/0000-0002-6354-3884}{Central (Pulkovo) Astronomical Observatory}{Laboratory of the Dynamics of the Galaxy}{Saint-Petersburg}{}{196140}{Russia}

\begin{abstract}
The recent results obtained by the modern telescopes and spacecrafts allow us for the first time
to compare directly the mass, spatial density and size distribution of the dust grains in the
regions of their production, processing and consumption in our Galaxy.
The ALMA and VLT/SPHERE telescopes allow us to estimate the production of the dust by supergiants
and collapsing core supernovae. The 2MASS, WISE, SDSS, Planck and other telescopes
allow us to estimate the processing of the dust in the interstellar medium.
After renewed Besan\c{c}on Galaxy model 
the medium appears to contain about half the local mass of matter (both baryonic and dark)
in the Galactic neighborhood of the Sun.
The Helios, Ulysses, Galileo, Cassini and New Horizons spacecrafts allow us to estimate the
consumption of the dust into large solid bodies.
The results are consistent each other assuming the local mean spatial density of the dust is about
of $3.5\times10^{-26}$ g/cm$^3$, mean density of the grain is about 1 g/cm$^3$,
and the dust production rate is about of 0.015 Solar mass per year for whole the Galaxy.
\end{abstract}

The dust grains are formed in the shells of branch red giants,
asymptotic branch giants, supergiants, novae and collapsing core supernovae
This is the production of the dust.
Then in the interstellar medium some events lead to the growth of the dust particles
whereas others lead to their fragmentation. 
This is the processing of the dust.
Finally, we meet the dust on the Earth as a result of the formation of the stars and planets from
the interstellar medium in past and the penetration of the dust particles into the Solar system
from the interstellar medium in present.
This is the consumption of the dust.

The production of dust by supergiants and collapsing
core supernovae are particularly important since only this production by very massive stars
can explain the large abundance of dust in high-z quasars.
For a long time, the observable mass of the dust production were several
orders of magnitude less than the theoretical estimates \citep{wesson}.
However, \citet{m2011}, \citet{m2015} and \citet{wesson} found the dust
mass of $0.5-0.8$ $M_{\odot}$ in the ejecta of SN~1987A by use of the Herschel Space Observatory.
\citet{inde} by use of ALMA telescope confirmed the condensation of more than 0.2 $M_{\odot}$
of carbon dust produced by SN~1987A for 25 years.
A noticeable fraction of the dust particles larger than 2 micrometers proves the intensive growth of
the grains in the ejecta.
Such large grains must better survive during the transfer to the interstellar medium.
Large grains with an average radius of 0.5 micrometers are also found by \citet{scicluna}
by use of the SPHERE instrument at the VLT telescope in the mass-loss envelope of the supergiant VY Canis Majoris.

\citet{li} estimated the mean rate of collapsing core supernova in the Galaxy as 2.3 per century.
If a typical supernova produces dust mass of 0.5 $M_{\odot}$, then the medium is enriched with
dust at a rate of 0.011 $M_{\odot}$ per year.
Together with supergiants it gives the total production about 0.015 $M_{\odot}$ per year.
At the Galactic star formation rate of about 1.6 $M_{\odot}$ a year \citep{licquia}
and the typical gas-to-dust ratio of 100, the star formation takes away from the medium about
of 0.016 $M_{\odot}$ a year.
Thus, the recent investigations proves the production and consumption balance.

For further comparison we use the power law distribution of dust particles by size in
the environment around the supernovae SN~2010jl, SN~1995N, SN~1998S, SN~2005ip and SN~2006jd
found by \citet{gall}.
Using the normalization of the dust spatial mass density of $3.5\times10^{-26}$ g/cm$^3$
discussed later 
this distribution is shown by the dash-dotted line in Figure 1 and discussed further as the
distribution of dust particles by size \textbf{in the places of their production}.

The mean density of the medium in the Galactic neighborhood within few hundred pc from the Sun
has revised significantly in the last years:
since 28\% of the total mass of matter (local spatial mass density 0.02 $M_{\odot}/pc^3$)
by the old version of the Besan\c{c}on model of the Galaxy \citep{robin} to
47\% (0.05 $M_{\odot}/pc^3$) by its new version \citep{czekaj} following 
\citet{binney}.
This increase of 2.5 times is explained by some discoveries of gas and dust far from the Galactic plane \citep{mckee}.
Thus, the interstellar medium contains about half the local mass of matter (baryonic and dark)
in the Galactic neighborhood of the Sun.
It looks reasonable if we allow the current star formation from the medium.
With the typical gas-to-dust ratio of 100 it means the local spatial mass density of dust
$5\cdot10^{-4}$ $M_{\odot}/pc^3$, or $3.5\times10^{-26}$ g/cm$^3$.
This value is used further.

The distribution of dust grains on their size has been investigated by many authors
(see review by \citet{g2016b}).
\citet{gobr}, \citet{planck}, \citet{daven}, \citet{g2012a, g2013a, g2013b, g2016a}
and other authors have used the observations with
Hipparcos (Tycho-2), 2MASS, SDSS, WISE, UKIDSS, Planck and other telescopes and
showed that the familiar medium with extinction-to-reddening coefficient $R_V=3.1$ is valid
only within 100 pc from the Galactic plane.
But for the bulk of the nearby kiloparsec and all investigated regions of the Galaxy up to 25 kpc
from the plane
the observations better agree with the dust model by \citet{wd2001} with $R_V=5.5$
or even with a flat extinction law in infrared, at $1-5$ micrometers.
It is supported by \citet{miville} showing that in a typical low-density
high-latitude clouds the relative velocities of typical grains of radius of about 0.1 micrometers
are close to the critical (growth vs. fragmentation) speed of order of 1 km/s.

In particular, \citet{g2016a} found an overdensity of large grains ($R_V=5.5$)
within 14 pc of the Solar system by use of the stellar photometry from the Tycho-2, 2MASS and WISE.
The orientation of this overdensity fits 16-year Ulysses spacecraft's dust
detector data about invasion of the dust and gas flow into the Solar system \citep{kruger}.
The Galactic coordinates of the entry point (about $l=3^{\circ}$, $b=21^{\circ}$)
of this flow roughly correspond to the area of maximum concentration of dust in the Gould belt
($l=15^{\circ}$, $b=19^{\circ}$) \citep{g2012b}.
Probably, there is a relation between the dust in Gould belt, in the Local overdensity and in the
Solar system.

I calculated the distribution of grains by size based on Mie theory \citep{bohren} and dust properties
derived for the Local overdensity by \citet{g2016a}.
The mean physical density of 3 g/cm$^{-3}$ was accepted for the grains as for silicates and graphite.
I obtained the maximum of size distribution at 0.25 micrometers and related
wavelength of the extinction maximum at 1.5 micrometers.
This distribution is shown in Figure 1 by the solid black curve.
It can be considered as the spatial mass distribution of grains by size
\textbf{in the interstellar medium}.

However, there are some reasons \citep{dwek}; \citep{bochkarev} to believe
that the mean physical density of grains is about 1 g/cm$^3$.
Some explanation is the models of fluffy, porous and/or icy grains \citep{zubko}.
With a fixed optical depth of the dust layer and a fixed spatial mass density of dust
the decrease of the physical density of grains means the increase of their size.
In the calculations based on Mie theory I found that to provide the extinction with the maximum
at $1.5$ micrometers the grains with radius of $0.25$ micrometers and physical density of 3 g/cm$^3$
can be replaced by the ones with radius of $1.5$ micrometers and physical density of 1 g/cm$^3$.
It means the shifted distribution shown in Figure 1 by black dotted line.
This curve can be considered as the spatial mass distribution of dust particles by size
\textbf{in the interstellar medium, more realistic}.

Based on the data of Ulysses (at 5 AU), Galileo (5 AU), Cassini (1--9.5 AU) and Helios (0.3--1 AU)
spacecrafts \citet{kruger} estimated the spatial density ($2.1\cdot10^{-27}$ g/cm$^3$)
and size distribution of \emph{interstellar} grains penetrating deep into the Solar system.
This distribution is shown in Figure 1 by the solid gray curve and is considered as
the one of the dust \textbf{inside the solar system}.
This estimate is an order of magnitude less than the one for interstellar medium.

According to \citet{poppe} the dust sensor aboard New Horizons spacecraft at
heliocentric distances $2.6-6.8$ AU showed the spatial density of \emph{interstellar} grains
close to that of Ulysses, Galileo, Pioneer 10, Pioneer 11, Voyager 1 and Voyager 2.
But for the distances $6.8-15.5$ AU New Horizons detected the mean spatial density of
$2.6\cdot10^{-26}$ g/cm$^3$ for \emph{interstellar} grains of $2\cdot10^{-12}-10^{-9}$ g
(i.e. a radius of $0.8-6.2$ microns).
This density is indicated in Figure 1 by the black square (horizontal line shows the uncertainty).
It can be considered as the assessment of dust \textbf{at the edge of the Solar system}.
Vertical shift in Figure 1 between the black square and gray curve reflect the slow decrease
of the interstellar dust flow deeper in the Solar system.

\section{Conclusions}

The results summarized in Figure 1 show that 
processing of the dust grains in the local interstellar medium shifts their distribution to smaller
grains,
the spatial mass density of dust in the medium is consistent with the one in the
outer regions of the Solar system,
only a small portion of dust penetrates deeper to the system.

\acknowledgements The study was financially supported by the
``Transient and Explosive Processes in Astrophysics''
Program P-7 of the Presidium of the Russian Academy of Sciences.

\articlefigure{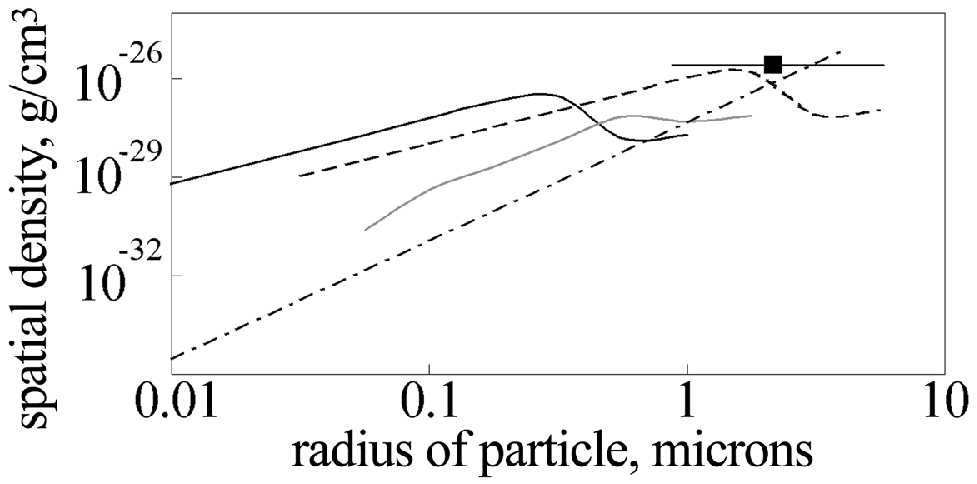}{g11}{The spatial mass density of dust vs. dust grain radius
in the supernova shell -- dash-dot line,
in the interstellar medium with grains of physical density of 3 g/cm$^3$ -- black solid line,
in the interstellar medium with grains of physical density of 1 g/cm$^3$ -- black dotted line,
in the inner Solar system -- gray solid line,
in the outer Solar system -- square with the line of uncertainty.}

\end{document}